# Binding Group of Oligonucleotides on TiO$_2$ Surfaces: Phosphate Anions or Nucleobases?

Federico A. Soria, Cristiana Di Valentin[*]

Dipartimento di Scienza dei Materiali, Università di Milano Bicocca,

via R. Cozzi 55 20125 Milano Italy

## Abstract

Although the immobilization of oligonucleotides (nucleic acid) on mineral surfaces is at the basis of different biotechnological applications, an atomistic understanding of the interaction of the nucleic acid components with the titanium dioxide surfaces has not yet been achieved. Here, the adsorption of the phosphate anion, of the four DNA bases (adenine, guanine, thymine, and cytosine) and of some entire nucleotides and dinucleotides on the TiO$_2$ anatase (101) surface is studied through dispersion-corrected hybrid density functional theory (DFT) calculations. Several adsorption configurations are identified for the separated entities (phosphate anion or base) and then considered when studying the adsorption of the entire nucleotides. The analysis shows that both the phosphate anion and each base may anchor the nucleotides to the surface in a collaborative and synergistic adsorption mode. The tendency is that the nucleotides containing the guanine base present the strongest adsorption while those made up with the thymine base have the lowest adsorption energies. Nucleotides based on adenine and cytosine have a similar intermediate behavior. Finally, we investigated the adsorption of competing water molecules to understand whether in the presence of the aqueous solvent, the nucleotides would remain bonded to the surface or desorb.

[*] Corresponding author: cristiana.divalentin@unimib.it

# 1 Introduction

Understanding the adsorption of biomolecules on mineral surfaces is a topic of fundamental relevance in the field of material science [1-3]. The applications are varied and include biomedicine,[4] drug delivery, [5,6] water treatment,[7] between others. Specifically, the immobilization of oligonucleotides (nucleic acids) on these surfaces constitutes the basis for different biotechnological applications [8,9], including genetic and infectious disease diagnostic devices [10] and miniaturized biosensor arrays [11].

Due to the molecular recognition properties through base pairing, nucleic acids are especially appealing for nanobiotechnology applications. The deoxyribonucleic acid (DNA) is composed of four nitrogenous bases: adenine (A), guanine (G), cytosine (C), and thymine (T). The first two are purines while the second two are pyrimidines. These bases can recognize each other through the complementary Watson−Crick base pairing [12], with A binding to T and C binding to G, selectively. Recognition and hybridization processes are crucial for DNA functioning in living organisms and for the detection, diagnostics and identification of genetic diseases and infectious agents. Moreover, these molecular recognition properties also permit programmable self-assembly of DNA strands into novel nanoscale structures.

Usually DNA is immobilized on gold, [13,14] glass, [15,16] oxidized silicon, [17,18] organic polymers [19,20]. The functionalization of $TiO_2$ surfaces nanoparticles (NPs) with DNA (single-stranded or double-stranded) was reported either through direct or indirect attachment. In the first case, the experimental results suggest that the extent of DNA adsorption depends on the pH: at neutral pH the binding is established mainly via the backbone phosphate but also the nucleobases can interact with the surface leading to a parallel adsorption configuration. However, the capacity of DNA adsorption is drastically increased at pH 2, since at this pH the bases do not interact with the surface because they are protonated leading to an upright adsorption configuration that allows a higher packing of DNA chains [21,22]. The indirect way to bind DNA to $TiO_2$ surfaces is through linkers. Levina et al. reported that DNA fragments can be covalently attached to polylysine linkers (positively charged), which are immobilized at the surface of $TiO_2$ nanoparticles (negatively charged) due to electrostatic interactions [23,24]. In the same line, the linking of DNA to $TiO_2$ NPs was achieved by using bridging enediol ligands, such as dopamine (DA), which facilitate photoinduced hole transfer across the bioinorganic interface [9,25].

From theoretical point of view, the interaction of biomolecules with $TiO_2$ surfaces was mostly investigated for aminoacids and peptides, interacting with flat $TiO_2$ surfaces and spherical nanoparticles [2,26]. Both, DFT geometry optimization and molecular dynamics (MD) simulations were reported in vacuum and in water environment. For instance, using DFT methods, Pantaleone et al. reported about the adsorption of canonical, zwitterionic, or deprotonated forms of different aminoacids on anatase (101) [27] or rutile (110) [28] surfaces. A larger reactivity of the rutile polymorph was observed both in adsorbing and in deprotonating the aminoacids, as compared with the anatase. Reactions of aminoacids polymerization, through the formation of peptide bonds, on the rutile (110) surface were studied at different levels of theory with reactive forcefield (ReaxFF), DFT and density-functional-based tight-binding (DFTB) methods [29]. From MD simulations, the free energies of adsorption were obtained for different aminoacid (polar, aromatic, charged and hydrophobic) on both flat surfaces and nanoparticles [30,31].

On the contrary, the interaction/adsorption of oligonucleotides (single or multiple) on $TiO_2$ surfaces has not yet been extensively studied from a computational point of view. In one study, the adsorption of the cytidine monophosphate nucleotide on the rutile (110) surface was investigated by DFTB calculations [32] and it was found that the nucleotide favors anchoring with two phosphate oxygen atoms [32]. In another one, the adsorption of the four DNA bases on the rutile (110) surface was studied through classical MD simulations and the potential of mean-constrained-force calculations. The simulations suggested that the nucleobases do not interact directly with the surface, but they adsorb on ordered water monolayers, which are in direct contact with the substrate [33].

In the present work, we provide a systematic study of the adsorption modes and energies for four DNA oligonucleotides on the anatase (101) surface by means of hybrid (B3LYP) density functional theory calculations. Grimme's approach in its D* formulation [34, 35] is used to consider the van der Waals interactions, which is well known to provide reliable descriptions of the structures and energetics of bio-molecules adsorption on mineral surfaces [36]. First, we explore the adsorption of the dihydrogen phosphate anion and of the four different bases on the surface, separately, considering different adsorption modes. Then, the adsorption of whole monomer of each type of nucleotide (phosphate anion + base) is studied. We observe a stable collaborative adsorption mode, where both the phosphate anion and the nucleobase are bonded to the surface. Further, we considered two different oligonucleotide dimers in different configurations on the surface. Finally, a comparative study, based on the adsorption energy of water, is presented. The atomistic insight allows us to understand in detail the surface-adsorbate interaction, usually not accessible experimentally, while comparison with the adsorption of water molecules give us a first assessment of the coating-desorption risk in an aqueous medium.

## 2 Computational Details

For all of the density functional theory (DFT) electronic structure calculations and geometry optimizations, we made use of the CRYSTAL14 simulation package [37], where the Kohn–Sham orbitals are expanded in Gaussian-type orbitals. The all-electron basis sets are Ti 86-411 (d41) and O 8-411 (d1) for $TiO_2$ and H 5-111 (p1), C 6-3111 (d1), O 8-4111 (d1), N 6-311 (d1) and P 85-211(d11) for the adsorbed oligonucleotide molecules. We used the B3LYP functional,[38,39] corrected by Grimme's D* to include dispersion forces [35]. The cutoff limits in the evaluation of the Coulomb and exchange series/sums were set to $10^{-7}$ for Coulomb overlap tolerance, $10^{-7}$ for Coulomb penetration tolerance, $10^{-7}$ for exchange overlap tolerance, $10^{-7}$ for exchange pseudo-overlap in the direct space, and $10^{-14}$ for exchange pseudo-overlap in the reciprocal space. For the geometry optimization, forces were relaxed to be less than $4.5 \times 10^{-4}$ au and displacements to be less than $1.8 \times 10^{-3}$ au.

To model the anatase (101) surface, three triatomic layers of $TiO_2$ slab were considered. The bottom layer was kept fixed to the optimized bulk positions during the geometry optimization (no periodic boundary conditions were imposed in the direction perpendicular to the surface). To investigate binding energies and equilibrium geometries, we used a $2 \times 4$ or $2 \times 6$ supercell model (144 or 216 atoms) which have eight or twelve $Ti_{5c}$ (5-fold coordinated) adsorption sites; a k-point mesh of $2 \times 2 \times 1$ to ensure the convergence of the electronic structure. The total adsorption energy per molecule has been defined as:

$$E_{ads} = (E_{slab+nmol} - [E_{slab} + n_{mol}E_{mol}])/n_{mol}$$

where $E_{slab+nmol}$ is the total energy of the whole system, $E_{slab}$ is the energy of the surface slab, $E_{mol}$ is the energy of the molecule in the gas phase, and nmol is the number of molecules adsorbed on the surface. The total density of states (DOS) and projected density of states (PDOS) were computed with a finer k-point mesh of $30 \times 30 \times 1$.

## 3 Results and Discussion

In order to rationalize the interaction of a single nucleotide with the anatase (101) surface, we decided first to study separately the adsorption of the two main components of the biomolecule, i.e. the dihydrogen phosphate anion, on one side, and the nucleobase, on the other. Thus, the paper is organized as follows: we present the adsorption of the dihydrogen phosphate group in **Section 3.1**, that of the four different nucleobases in **Section 3.2**, whereas in **Section 3.3** we analyze the adsorption of the four complete nucleotides, in different adsorption modes. In **Section 3.4** we compare the adsorption energies of the dinucleotides on the surface. Finally, in **Section 3.5**, a comparative study with water adsorption is presented.

### 3.1 Dihydrogen Phosphate Adsorption

The chemically active sites of the anatase $TiO_2$ (101) surface are the 5-fold coordinated cationic Ti atoms ($Ti_{5c}$) and the 2-fold coordinated anionic O atoms ($O_{2c}$). There are different possibilities in which dihydrogen phosphate anion can bind to a metal-oxide surface through bonds between phosphate oxygen atoms and surface metal atoms. The coordination can be either monodentate or bidentate, depending on how many of the phosphate oxygen atoms bind to the surface Ti atoms. Also, the possibility of O-H dissociation and of the H migration to the different surface $O_{2c}$ must be considered.

*3.1.1. Monodented adsorption mode*

A molecular monodented adsorption takes place when one dehydrogenated oxygen atom of the dihydrogen phosphate anion binds to a surface $Ti_{5c}$ atom, with a possibility for the two hydroxyl groups to establish hydrogen bonds (H-bonds) with two surface $O_{2c}$ atoms. Two monodented adsorption modes could be localized, which are denoted as M1 and M2 in **Figure 1.** In the M1 configuration, only the covalent bond (1.92 Å) between one O atom of the dihydrogen phosphate anion and a surface $Ti_{5c}$ atom is formed, with a calculated adsorption energy of -0.86 eV. The other most stable monodented adsorption mode M2 involves the formation of one O-$Ti_{5c}$ bond (1.86 Å) and one H-bond with a $O_{2c}$ atom of the surface (1.81 Å), with an adsorption energy of -1.02eV.

*3.1.2. Bidented adsorption mode*

In the bidented structures, two O atoms of the $H_2PO_4^-$ anion bind to two surface $Ti_{5c}$ atoms. Two configurations, B1 and B2 in **Figure 1**, were localized. The B1 configuration results from one proton dissociation of the dihydrogen phosphate and its migration to a surface $O_{2c}$ atom. In the B2 structure, no deprotonation is observed since the two phosphate O atoms bind to the surface Ti atoms. The adsorption energies are -1.66 and -2.01 eV for B1 and B2, respectively. This proves a larger stability of the undissociated mode on the surface. This behavior of the dihydrogen phosphate ion is different to what observed in the case of phosphonic or formic acids, where dissociated adsorption was found to be preferred [40,41], but it is expected based on the acidic character of the compared species: while the pk$_a$ of phosphonic and formic acids are as low as 1.3 and 3.75, respectively, for the dihydrogen phosphate anion it is as high as 7.20. This explains the lower tendency of the $H_2PO_4^-$ anion to dissociate in comparison with phosphonic or formic acids.

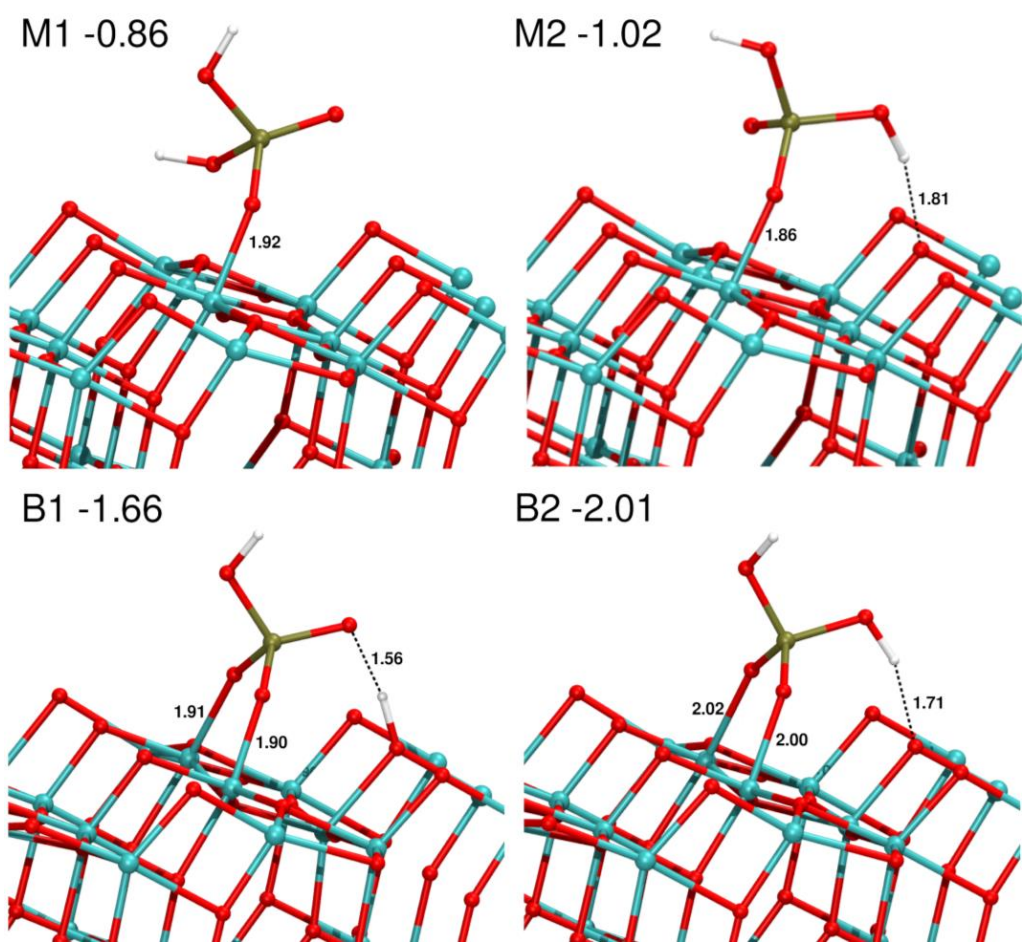

**Figure 1.** Adsorption modes and adsorption energies per molecule (in eV) of dihydrogen phosphate adsorbed on the TiO$_2$ (101) anatase surface as obtained by B3LYP-D*. Top, monodented modes. Bottom, bidented modes. Distances in Å.

From the adsorption energies for the different modes, bidented undissociated adsorption mode being preferred. This result has important implications for the adsorption of single nucleotides on the anatase (101) surface. A single nucleotide has two oxygen atoms available in the phosphate group to attach to the TiO$_2$ surface and one hydroxyl group free to establish one H-bond with one surface O$_{2c}$ atom. In the case of longer ssDNA chains, other considerations must be taken in account, such as the conformational flexibility of the chain [42,43].

*3.1.3. Electronic Structure*

It is interesting to analyze the electronic structure of these systems. B3LYP functional, given the portion of exact exchange included, provides a better framework given that the band gap of TiO$_2$ is largely underestimated by standard DFT methods. The main question is whether the dihydrogen phosphate anion induces the formation of new states in the band gap of the material, which could have interesting consequences, such as the extension of the absorption edge into the visible spectrum. From the total and projected density of states (DOS) reported in **Figure 2** for the different structures considered, one can observe that the introduction of additional midgap states arising from the presence of the adsorbed H$_2$PO$_4^-$ anion depends on the adsorption mode, whereas the band gap of the material ($E_g$) is essentially not affected. For instance, for the bidented B2 adsorption mode, the molecular states fall in the same energy range of the valence band and higher in energy with respect to the lower part of the conduction band, thus the band gap is still empty. On the contrary, for the bidented B1 as well as for the monodented M1 and M2 adsorption modes, new states appear in the band gap of TiO$_2$. These results suggest a sensible red shift of the optical band gap of the TiO$_2$ surface resulting from the dihydrogen phosphate adsorption.

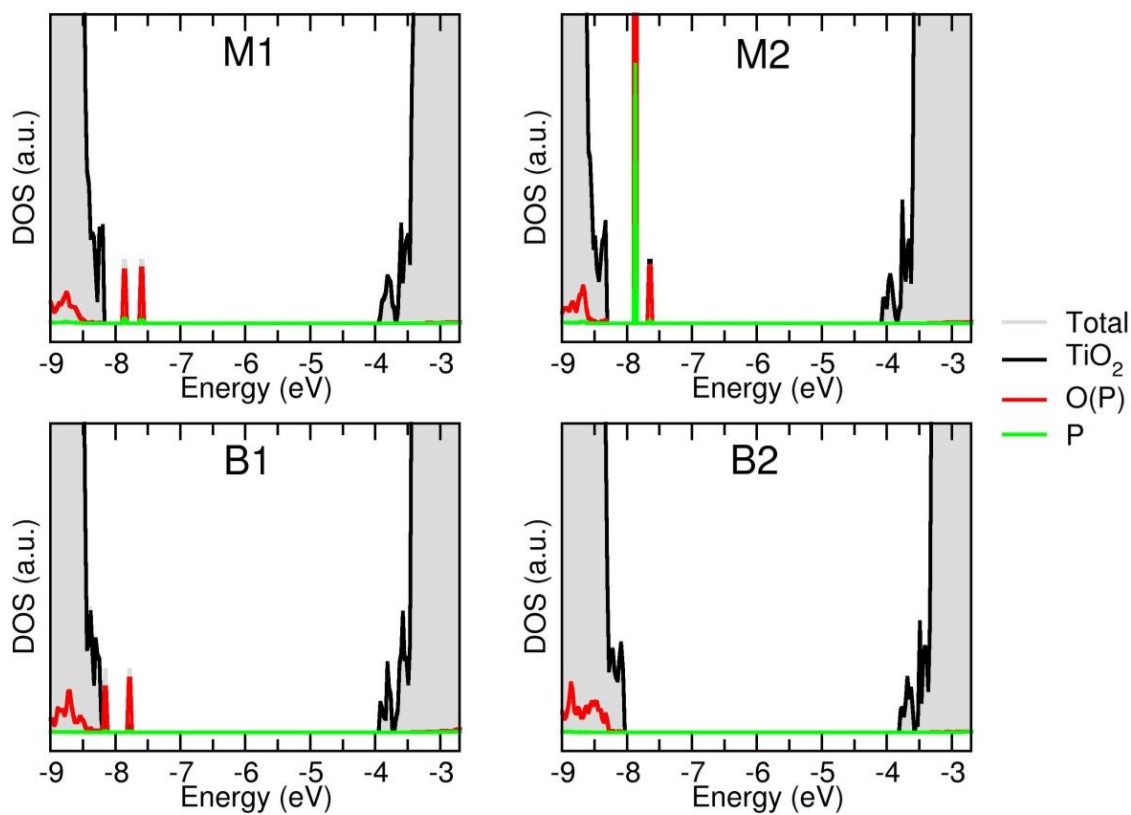

**Figure 2.** Total (DOS) and projected (PDOS) density of states for the adsorption of dihydrogen phosphate on the (101) anatase $TiO_2$ surface.

### 3.2 Nucleobases Adsorption

In this section, we study the adsorption of the four different nucleobases on the (101) $TiO_2$ anatase surface. The aim of this part is to investigate the interaction of this component of a nucleotide with the surface. Considering the different chemical nature of the four nucleobases, one may conclude that each present different possible interacting species: in all nucleobases there are different aromatic N atoms, moreover, in adenine, guanine and cytosine there are also amine groups (-$NH_2$), and in guanine, cytosine and thymine there is, at least, one carbonyl oxygen atom, which can interact with the $Ti_{5c}$ on the surface.

*3.2.1. Purines adsorption*

In **Figure 3** we report the optimized structures with some equilibrium distances when adenine and guanine nucleobases are adsorbed on the anatase $TiO_2$ (101) surface. In the case of adenine, we found three stable configurations. In the first, we observe the interaction of one aromatic N with a $Ti_{5C}$ surface atom (2.37 Å) and the H-bond between one H atom of the $NH_2$ group and a surface $O_{2c}$ atom (see **Figure**

**3a)**. The calculated E$_{ads}$ is -1.76 eV. In **Figure 3b** a second more stable structure (E$_{ads}$=-1.94 eV) is shown, where one aromatic N atom (opposite to the NH$_2$ group) is bound to a Ti$_{5C}$ surface atom with a bond distance of 2.30 Å. Finally, the third configuration (**Figure 3c**) is characterized by the largest adsorption energy of -2.19 eV and presents two simultaneous N-Ti$_{5C}$ bonds: one involving the N atom of the amine group (N-Ti distance 2.41 Å) and the other involving an aromatic N atom (N-Ti distance 2.40 Å).

In the case of guanine, three adsorption modes were also identified. In **Figure 3d** we present the structure where three H-bonds are established together with a weak interaction between an aromatic N atom and a Ti$_{5c}$, with E$_{ads}$=-1.45 eV. In **Figure 3e** the adsorption configuration displays an interaction between the N atom of the amine group and a Ti$_{5c}$ atom together with a H-bond between the H atom of the NH in the six-member ring and an O$_{2c}$ with an adsorption energy of -1.67 eV. However, the strongest interaction (-2.42 eV) is observed when the carbonyl oxygen of the guanine forms a covalent bond with a Ti$_{5c}$ of the surface (**Figure 3f**) together with an aromatic N atom bound to a surface Ti$_{5c}$ atom.

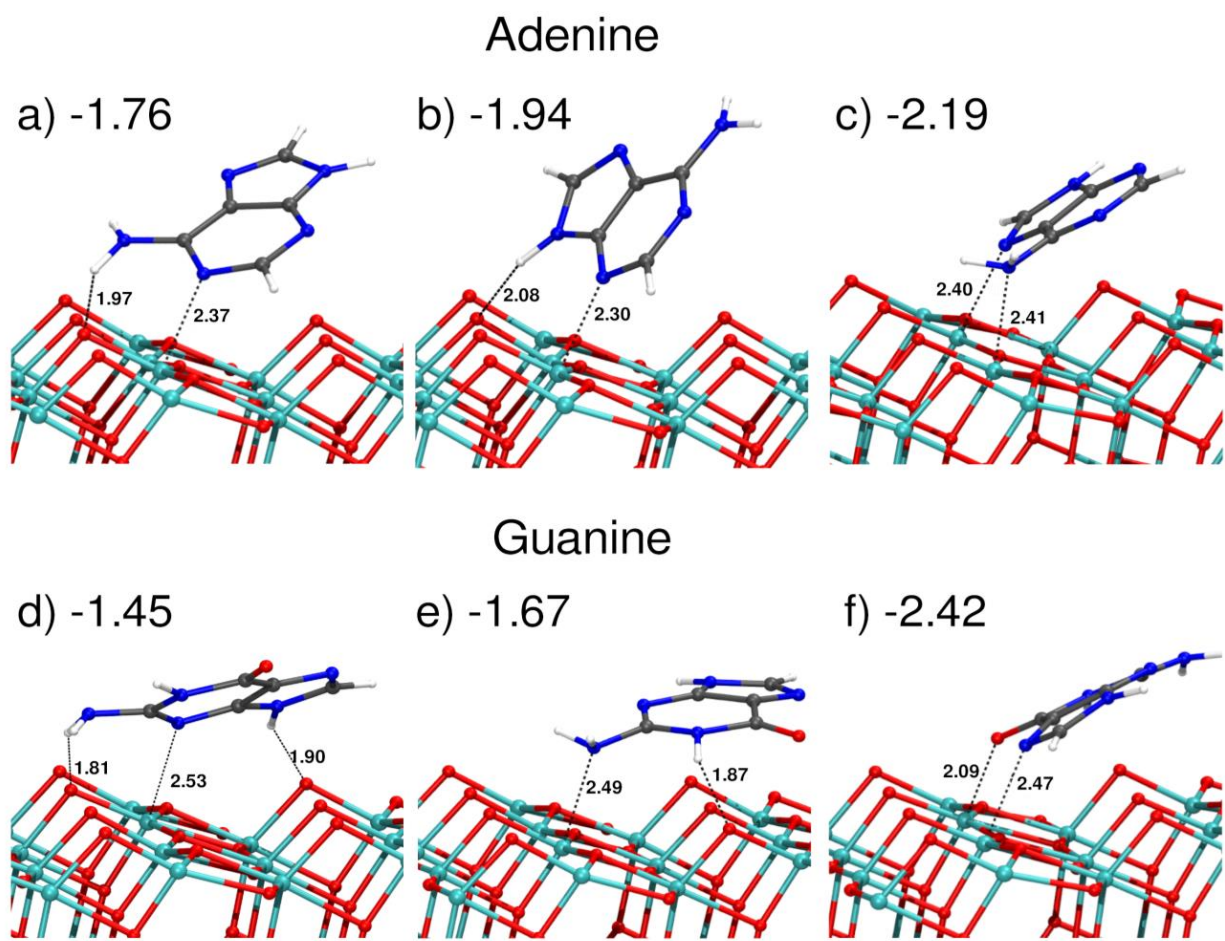

**Figure 3.** Adsorption modes and adsorption energies per molecule (in eV) of purine nucleobases adsorbed on the TiO$_2$ (101) anatase surface as obtained by B3LYP-D*. Top, adenine nucleobase. Bottom, guanine nucleobase. Distances in Å.

*3.2.2. Pyrimidines adsorption*

As for purines, also in the case of pyrimidines we found different adsorption modes on the surface. For cytosine, we identified three adsorption modes: the first one, with an adsorption energy of -1.78 eV, presents an interaction of an aromatic N atom with a Ti$_{5c}$ atom on the surface accompanied by a H-bond between the amine group and a surface O$_{2c}$ atom (**Figure 4a**). In the second one (**Figure 4b**), with an adsorption energy of -1.88 eV, a bond between the carbonyl O atom of cytosine and a surface Ti$_{5C}$ (2.09 Å) together with a H-bond between an amine group and a surface O$_{2c}$ is observed. Finally, a higher adsorption energy of -1.96 eV is obtained for the third structure (**Figure 4c**), when the carbonyl oxygen and the aromatic N atom are bound to two different Ti$_{5c}$ atoms with bond lengths of 2.35 Å and 2.53 Å, respectively.

In the case of thymine, lower adsorption energy values are computed. Since this nucleobase lacks amine groups and the ring N atoms are hydrogenated, the two carbonyls O atoms can be mostly involved in the interaction with the surface. The difference between the various adsorption modes is related to the number of O-Ti$_{5c}$ bonds, to the formation of extra H-bonds or to the orientation of the ring plane with respect to the surface. For the structure in **Figure 4d**, no H-bonds are established, and the ring is perpendicular to the surface (E$_{ads}$=-0.84 eV). In **Figure 4e**, both carbonyl O atoms bind to two different surface Ti$_{5c}$ and the ring plane is parallel to the surface favoring the van der Waals interactions (E$_{ads}$=-1.53 eV). Finally, the strongest interaction (-1.66 eV) is observed when one O-Ti$_{5c}$ bound is accompanied by the formation of a H-bond between one NH group and a surface O$_{2c}$ atom (**Figure 4f**).

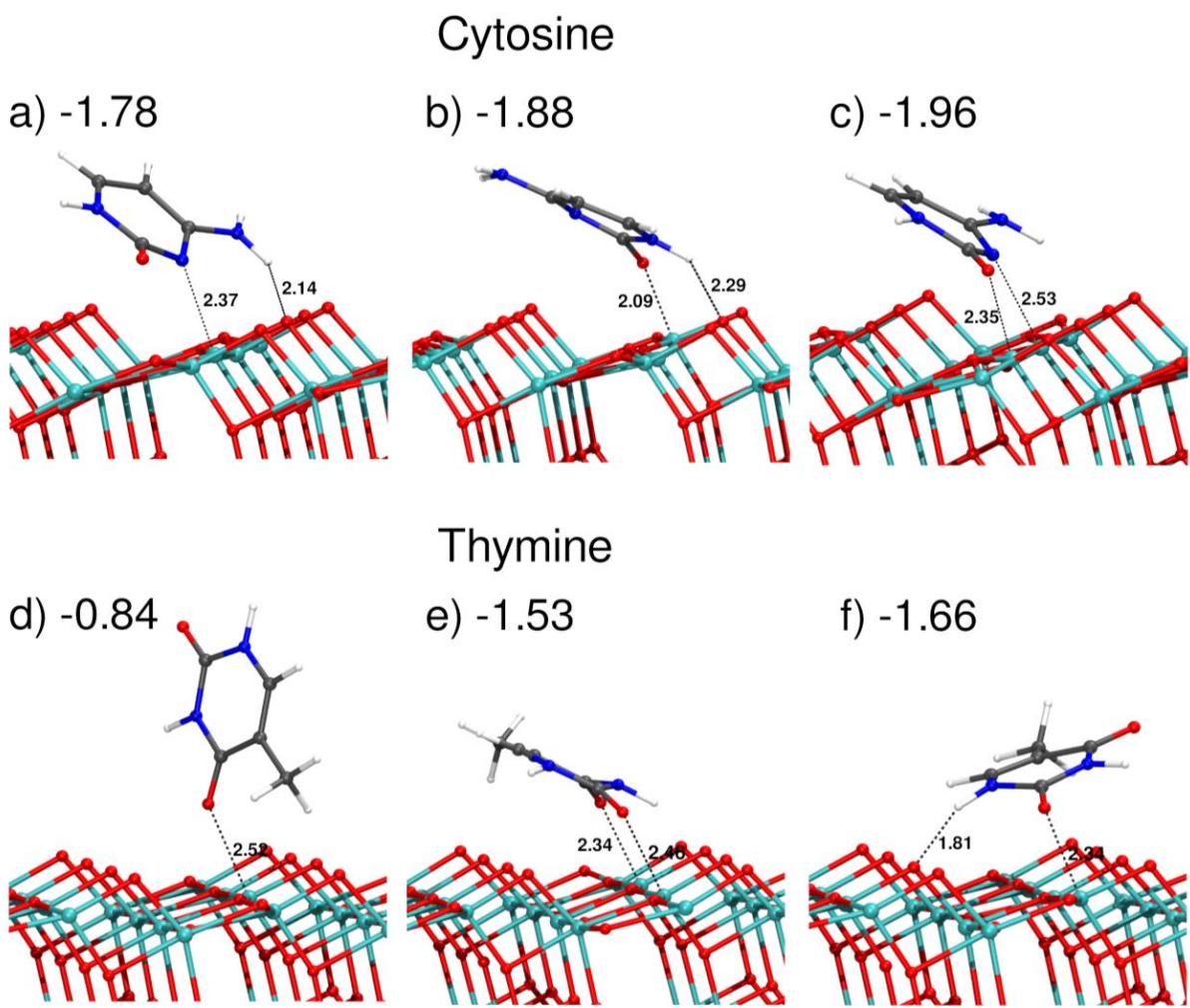

**Figure 4.** Adsorption modes and adsorption energies per molecule (in eV) of pyrimidine nucleobases adsorbed on the TiO$_2$ (101) anatase surface as obtained by B3LYP-D*. Top, cytosine nucleobase. Bottom, thymine nucleobase. Distances in Å.

Considering the adsorption energies for all three nucleobases in all the adsorption modes it can be observed that for purines nucleobases and for cytosine the adsorption energies are between -1.5 and -2.4 eV, while for thymine the range is between -0.8 and -1.6 eV. Taking the most stable structure of each nucleobase, the adsorption strength is in the order $T < C < A < G$, which is in line with what observed for other surfaces, such as graphene or Au(111) [44 - 46].

Comparing with other surfaces, on anatase TiO$_2$ (101) the nucleobases adsorb with higher adsorption energies. For instance, on graphene and hexagonal boron nitride (*h*-BN) the binding energies are in the range of 0.93–1.18 eV, and the adsorption is principally due to the vdW interactions between the nucleobases and the atomic monolayer [44,45]. On Au (111) surface, Rosa et al. reported, based on DFT calculations with the vdW–DF functional, that the preferential adsorption mode for these nucleobases is parallel to the surface with adsorption energies in the range between 0.7 and 1.0 eV [46]. The differences can be attributed to the fact that on the anatase surface the adsorption of nucleobases involve covalent bonds, vdW interactions and H-bonds.

In general, we found that the types of interaction and the adsorption mode of nucleobases on $TiO_2$ are similar that those reported for other metal oxides or minerals surfaces. For instance, on ZnO [47] and $SiO_2$ [48] clusters the adsorption of guanine was found to be characterized by a strongly covalent character due to the interaction between an aromatic N or a carbonyl O atom of the guanine and a Zn (or Si, respectively) atom of the cluster together with the formation of a H-bond between the $NH_2$ group and one O atom on the ZnO or $SiO_2$ cluster. As we discussed above, similar types of interactions are observed for the adsorption of nucleobases on anatase (101) $TiO_2$ surface. Regarding to the adsorption mode, the binding of the nucleobases on montmorillonite was found to be stronger in a surface-parallel configuration (in analogy to what we found in this study) than in the orthogonal ones, due to a better balance between electrostatic and dispersion contributions [49]. The interaction energies on montmorillonite are around -1.8 and -2.2 eV, similar to the values that we have calculated in this work for $TiO_2$. In **Figures 3** and **4** it is possible to see that the nucleobases are generally parallel or slightly inclined of a low angle with respect to the surface.

### 3.3 Single Nucleotide Adsorption

According to the results in the previous sections, the adsorption energies of the phosphate anion, depending on the adsorption mode, are between -1.0 and -2.0 eV. Thus, one can expect that for a complete nucleotide (i.e. hydrogen phosphate anion + deoxyribose sugar + nucleobase) a collaborative adsorption mode will be present, where both the hydrogen phosphate anion and the nucleobase interact with the surface. In this section, we will show how this collaborative adsorption mode increases the adsorption energy of the nucleotide, with respect to the isolated portions of the nucleobase and of the phosphate anion.

The nucleotides are named deoxyadenosine monophosphate (dAMP), deoxyguanosine monophosphate (dGMP), deoxycytidine monophosphate (dCMP) and deoxythymidine monophosphate (dTMP). Based on the adsorption energies of the dihydrogen phosphate anion or the different nucleobases calculated in the previous sections, it possible expect a collaborative adsorption mode involving both the nitrogenous bases and the phosphate anion. In **Figure 5** the optimized structures on the anatase (101) surface are displayed with equilibrium distances. From top to bottom the order is: dAMP, dGMP, dCMP and dTMP. We investigate three different binding modes for the phosphate anion to the surface: monodented, bidented dissociated or bidented undissociated. It is worth noting that the bond distances of the O atom of the phosphate with the surface $Ti_{5c}$ atoms are slightly higher when we go from the monodented to the bidented undissociated adsorption mode, as we observed for the isolated dihydrogen phosphate anion (in **Section 3.1**). We observe that, independently of the phosphate anion binding mode to the surface, the nucleobase bends to the surface in order to form either H-bonds with the surface $O_{2c}$ atoms or to interact with surface $Ti_{5c}$ atoms through the N or O atoms of the nucleobase

In **Table 1** the adsorption energies for each configuration of the four nucleotides on the surface are reported. In the case of dAMP, the adsorption energies are -2.61, -3.64 and -3.89 eV for monodented, bidented dissociated and bidented undissociated, respectively, which are larger than those computed for the dihydrogen phosphate anion (**Figure 1**), on one side, and for the adenine base (**Figure 2**), on the other. This suggests a collaborative adsorption between the adenine base and the phosphate anion on the surface. The same behavior is observed for all four nucleotides.

For all four nucleotides, the bidented undissociated adsorption mode is the most stable. The adsorption strength follows the same trend as that for the isolated nucleobases, i.e. dTMP < dCMP < dAMP < dGMP. This also shows that the adsorption of the nucleotide is mostly determined by the type of nucleobase contained.

In order to confirm the presence of two simultaneous interactions (through to the phosphate anion and the nucleobase) we analyzed the charge density difference, $\Delta\rho(r)$, for dGMP in the bidented undissociated mode, to gain more insight into the chemistry of the monomer anchoring the $TiO_2$ (101) surface, as shown in **Figure S1**. In the $\Delta\rho(r)$ plots, the red color represents electron charge accumulation, and the blue color represents electron depletion regions with respect to the separated fragments ($TiO_2$ surface and monomer), in the same atomic positions as when in the complex. The charge density difference shows a clear increase in electronic density along the O-Ti bonds of the phosphate and an accumulation along the bond of the carbonyl O atom of the guanine nucleobase and the $Ti_{5c}$ surface atom.

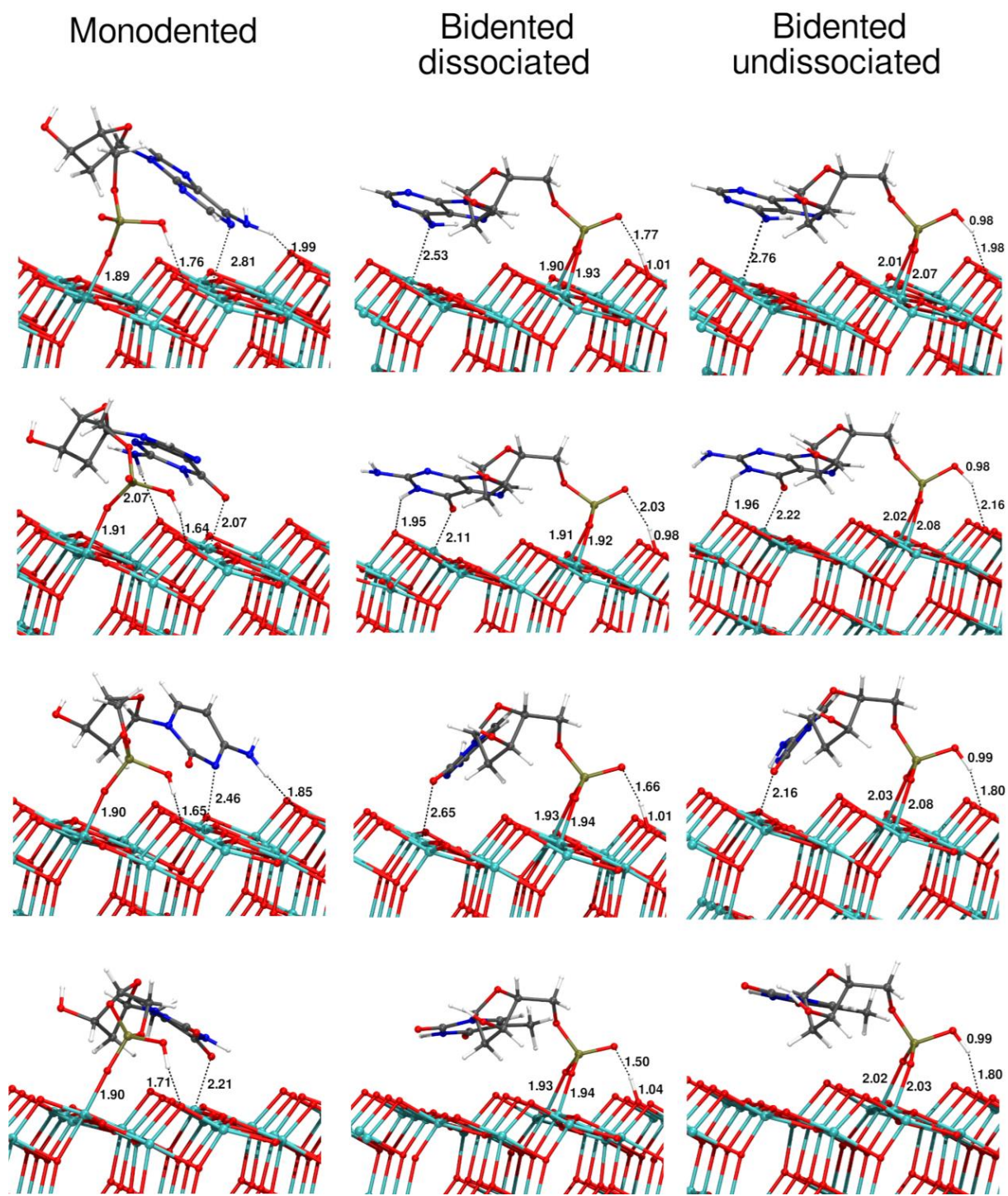

**Figure 5.** Optimized structures of the adsorption of nucleotides on the TiO$_2$ (101) anatase surface in monodented, bidented dissociated and bidented undissociated mode as obtained by B3LYP-D*. From top to bottom: dAMP, dGMP, dCMP and dTMP. Distances in Å.

**Table 1:** Total adsorption energy (E$_{ads}$), (expressed in eV) for the single nucleotides adsorbed on TiO$_2$ (101) anatase surface for the optimized structures reported in the **Figures 5**.

|      | E$_{ads}$(eV) |       |       |
| ---- | ----- | ----- | ----- |
|      | M     | B-D   | B-UD  |
| dAMP | -2.61 | -3.64 | -3.89 |
| dGMP | -3.37 | -4.13 | -4.44 |
| dCMP | -2.62 | -3.46 | -3.97 |
| dTMP | -2.04 | -2.85 | -2.87 |
|  |  |  |  |
|  |  |  |  |
|  |  |  |  |
|  |  |  |  |
|  |  |  |  |

Our findings on TiO$_2$ are in very good agreement with previous results on Al$_2$O$_3$ surfaces. In particular, the adsorption of dTMP on this oxide was investigated through the combination of first-principles calculations, X-ray photoelectron spectroscopy (XPS) and Fourier transform infrared (FTIR) spectroscopy [50]. The experiments demonstrate that dTMP covalently binds to the alumina surfaces. Calculations indicate that the covalent bonding of all the dTMP polar groups (sugar ring, phosphate group, and thymine) is thermodynamically favored. However, spectroscopic data and theory-based assignments of vibrational modes show that the bonding takes place primarily through the thymine and phosphate groups [50]

### 3.4 Dinucleotide Adsorption

In order to make our model closer to a realistic situation, in this section we investigate adsorption of a dinucleotide as a prototype system for an oligonucleotide, which is commonly used to functionalize NP surfaces. A dinucleotide is a short single stranded DNA formed by two monomers. For each dinucleotide, we studied two adsorption modes: the first involves the binding of two phosphate anions and the nucleobases to the surface, whereas the second involves the binding of a phosphate and a nucleobase to the surface. Since in the case of monomers, we found out that dGMP and dTMP are the nucleotide with the strongest and the weakest binding energy, respectively, in **Figure 6** we show the structures of the dinucleotides containing guanine or thymine nucleobases adsorbed on the anatase (101) surface, while in **Figure S2** we present the complete set of results where all nucleobases are considered.

In **Figure 6** (upper left) we present the optimized structure of a dimer of dGMP adsorbed on the surface when one phosphate unit and one nucleobase bind to $Ti_{5c}$ ions (namely 2×dGMP-UP). In order to keep the pi stacking conformation between the guanine residues, the monodentate adsorption is preferred. The binding mode is similar to what observed for a single nucleotide (**Figure 5**) because the second monomer is not interacting with the surface. The adsorption energy for this configuration is -2.63 eV. In **Figure 6** (upper right) the adsorption structure of the same dimer through the binding of two phosphate anions and two bases (2×dGMP-DOWN) is shown. For a comparison on an equal footing with the first adsorption configuration, we selected the monodentate phosphate binding mode. In this configuration the calculated adsorption energy is much larger: $E_{ads}$= -5.46 eV. This shows that the dinucleotide prefers the DOWN against the UP configuration. In the case of the dinucleotide, it is possible to define an adsorption energy *per* interacting nucleobases or $E^N_{ads}$. In the UP configuration since just one nucleobase is interacting with the surface $E^N_{ads} = E_{ads}$= -2.63 eV whereas in the DOWN configuration, since the two bases interact with the surface, $E^N_{ads}$ is half of the adsorption energy: $E^N_{ads}$= -2.73 eV. We conclude that the normalized adsorption energies for the two configurations are very close one to the other (-2.63 vs. -2.73 eV), indicating that the base/surface interaction is similar in both structures.

**Figure 6.** Optimized structures of the adsorption of a dimer of dGMP (top) and dTMP (bottom) on the TiO$_2$ (101) anatase surface in two different adsorption modes (see text) as obtained by B3LYP-D*. Distances in Å.

In the case of a dTMP dimer, we found that the adsorption through one phosphate anion and one nucleobase (2×dTMP-UP) has an adsorption energy of -2.25 eV (**Figure 6** (lower left)). In this configuration, the other nucleobase of the dimer does not interact with the surface, similarly to the 2×dGMP-UP. Compared with the 2×dGMP-DOWN configuration, a lower adsorption energy is observed (-2.63 eV vs. -2.25 eV). This result is in line with what computed for the single monomer in the previous paragraph, where the guanine base interacts more strongly with the surface than thymine, even if the same chemical group (carbonyl O atom) is involved. In **Figure 6** (lower right) we present the optimized structure for the adsorption of 2×dTMP in DOWN configuration. For this configuration the adsorption energy is $E_{ads}$= -4.60 eV. Again, a large difference in the adsorption energy as compared with dGMP is observed (-5.46 eV vs -4.60 eV), which confirms that the energy difference is due to the different nucleobases. When considering the adsorption energy *per* interacting nucleobase, $E^N_{ads}$, we observe a similar behavior as in the 2×dGMP: UP and DOWN configurations present similar values (-2.25 vs. -2.30 eV, respectively) suggesting a similar base/surface interaction.

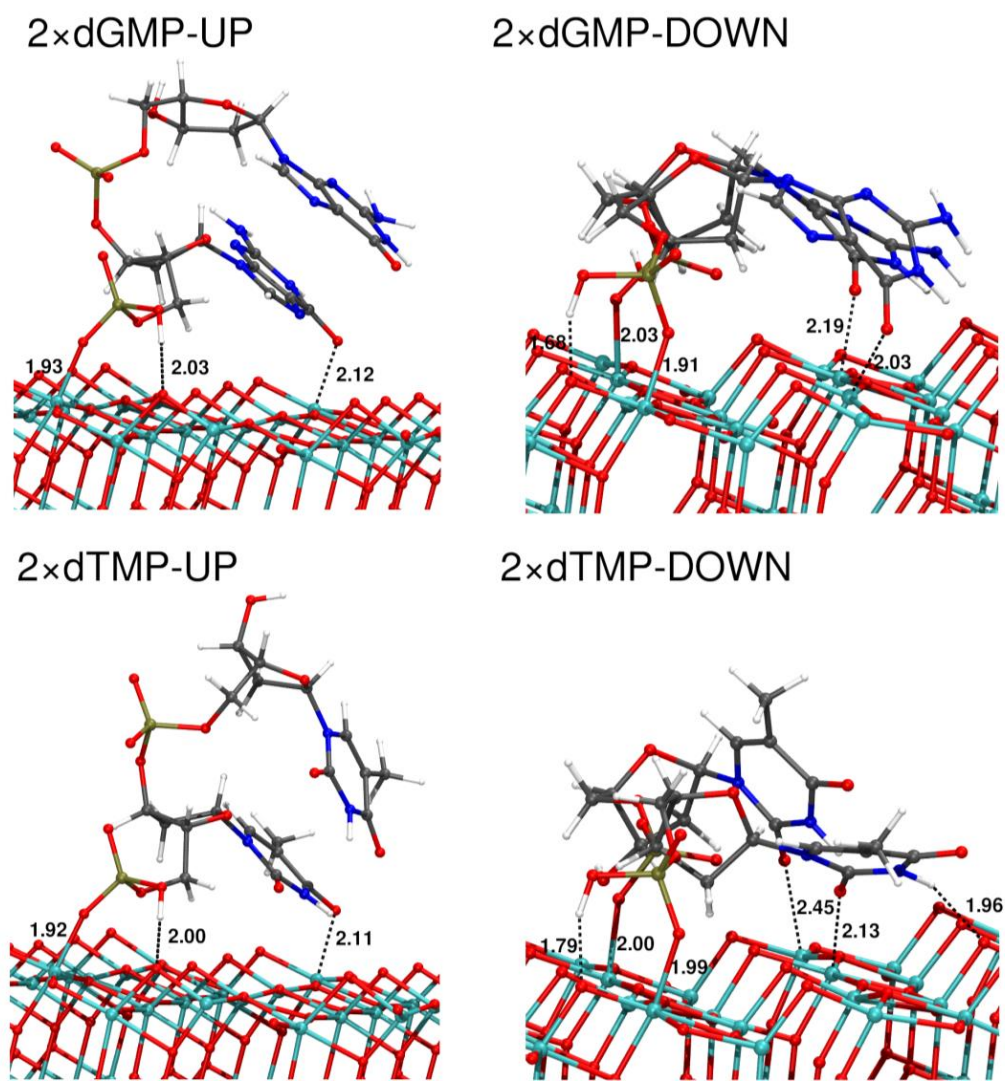

In **Table 2** we present the adsorption energy values for all types of dinucleotide in both UP and DOWN configurations. The DOWN configuration is clearly always preferred over the UP one. In the DOWN configuration, the guanine-based dimer anchors to the surface with the largest binding energy, followed by the 2×dAMP and 2×dCMP with similar adsorption energies and, last, the thymine-based dimer with the lowest binding energy. Differently, in the UP configuration, 2×dGMP is characterized by the highest adsorption energy, whereas the other three dimers present similar binding energy values.

**Table 2:** Total adsorption energy ($E_{ads}$), (expressed in eV) for the dinucleotides adsorbed on TiO$_2$ (101) anatase surface.

|  | $E_{ads}$ | |
|---|---|---|
|  | UP | DOWN |
| 2×dAMP | -2.40 | -5.29 |
| 2×dGMP | -2.63 | -5.46 |
| 2×dCMP | -2.39 | -5.32 |
| 2×dTMP | -2.25 | -4.60 |

It is important to note that, even though the DOWN configuration, where both phosphate groups and bases bind to the surface, is energetically favored, in a real ssDNA oligonucleotide, with a higher number of nucleotide units, other factors, such as the conformational disposition of the ssDNA, may come into play, which could oppose to the simultaneous interaction of phosphate groups and bases with the surface.

The adsorption of oligonucleotides, labeled with FAM (6-carboxyfluorescein), on TiO$_2$ anatase nanoparticles was experimentally studied by Zhang et al. [21] They reported that at pH=2 the adsorption capacity is much higher (ten times) than that at pH=7. They suggested that the electrostatic interactions are responsible for the attachment of the oligonucleotide chain to the nanoparticle. At pH=2 the nanoparticle's surface is expected to be positively charged, while the ssDNA fragments present negative charges due to the backbone phosphate anions. The nucleobases do not or only weakly interact with the surface since adenine and cytosine bases are protonated at pH=2, while guanine and thymine are partially protonated or non-protonated, respectively. For these reasons, an upright conformation of ssDNA on the NP is proposed [21]. On the contrary, at neutral pH, the adsorption capacity (it means the number of oligonucleotides bonded to the surface) decreases suggesting that the ssDNA wraps up around the $TiO_2$ NPs [21] These experimental observations agree with our computational results: indeed nucleotide or dinucleotide (when they are not protonated as in our models) prefer a collaborative adsorption through both the phosphate anions and the nucleobases leading to a down or parallel (to the surface) adsorption mode that implies a larger contact with the $TiO_2$ surface.

It is worth noting that this is possible for a single strand DNA. For a double stranded DNA, since the nucleobases are interacting with each other through complementary Watson−Crick base pairing, their possible interaction with the surface is forbidden and the phosphate anions are the main binding species to link DNA to the surface.

### 3.4.1. Electronic Structure

In this section, we analyze and compare the electronic structure for the 2×dGMP and 2×dTMP dimers adsorbed on the anatase (101) surface in terms of the total (DOS) and projected density of states (PDOS). The electronic structures for the different adsorption configurations are shown in **Figure 7**. We report in **Table 3** the values for the electronic ($E_g$) and HOMO–LUMO ($\Delta_{HOMO-LUMO}$) band gaps for the four structures under investigation.

The PDOS of 2×dGMP-UP structure (**Figure 7**) presents nine midgap states. The HOMO is predominantly constituted by one π state of the guanine base, which is far from the surface, whereas the HOMO–1 is localized on the guanine base close to the surface. The HOMO–2 and HOMO–3 states are mainly constituted by the π states of the guanine base far from the surface. The rest of the states are made up of a mixture of states from the bases, the sugar groups and the phosphate anions. A similar behavior is observed for the 2×dGMP-DOWN structure. Eight midgaps are present in the electronic structures. From HOMO to HOMO–3 states are constituted by the π states of the guanine bases. The midgap states close to the valence band of the $TiO_2$ are associated to a mixture of states coming from the bases, the sugar groups and the phosphate anions. The value of $\Delta_{HOMO-LUMO}$ is larger for 2×dGMP-DOWN than for 2×dGMP-UP **(Table 3)**.

**Table 3:** Electronic ($E_g$) and HOMO−LUMO ($\Delta_{HOMO-LUMO}$) band gaps in eV for the dimers of dGMP and dTMP adsorbed on the TiO$_2$ (101) anatase surface according to the optimized structures reported in the **Figures 7**.

|  | $E_g$ | $\Delta_{HOMO-LUMO}$ |
|---|---|---|
| 2×dGMP-UP | 4.20 | 2.19 |
| 2×dGMP-DOWN | 4.22 | 2.32 |
| 2×dTMP-UP | 4.21 | 2.46 |
| 2×dTMP-DOWN | 4.26 | 2.87 |

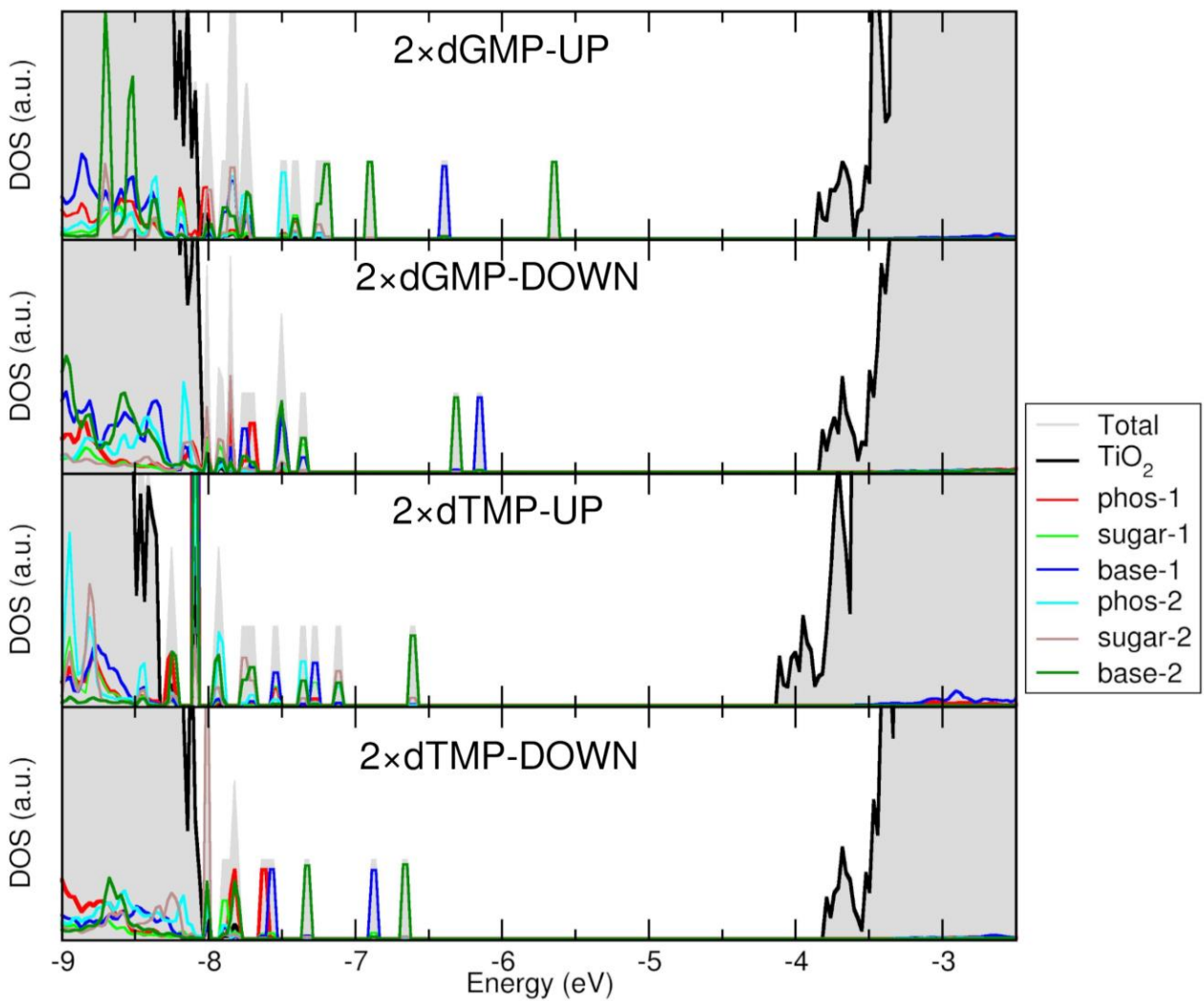

**Figure 7.** Total (DOS) and projected (PDOS) density of states for the adsorption of a dimer of 2×dGMP and 2×dTMP on the TiO$_2$ (101) anatase surface in two different adsorption mode (see text) as obtained by B3LYP-D*.

The electronic structure of adsorbed 2×dTMP is similar to that of adsorbed 2×dGMP: for the 2×dTMP-UP configuration nine midgap states are present. The HOMO state is formed by the π states of the thymine base far from the surface, while from HOMO–1 to HOMO–8 states are made up of a mixture of states from the bases, the sugar groups and the phosphate anions. Conversely, for the 2×dTMP-DOWN structure, six midgap are present. In this case the states from the HOMO to the HOMO-3 are associated to both thymine bases while the rest of the midgaps are a mix of the states coming from the bases, the sugar groups and the phosphate anions. As in the case of the 2×dGMP, the $\Delta_{HOMO-LUMO}$ values are larger for 2×dTMP-DOWN than for 2×dTMP-UP **(Table 3).**

|  |  |  |
|--|--|--|
|  |  |  |
|  |  |  |
|  |  |  |
|  |  |  |

## 3.5 Comparison between Nucleotides (monomers and dimers), and Water Adsorption.

In this section, in order to provide a first assessment on the thermodynamic competition for adsorption between the oligonucleotides and water, the most common solvent in the human body, we compare the adsorption energy of molecular water, which is well known to be the most stable, at a coverage of 0.25 and 1.0 ML on the anatase $TiO_2$ (101) surface.

The details of water adsorption on the anatase $TiO_2$ (101) surface are still controversial. From both the theoretical and the experimental point of view, it is not yet clear what is the precise extent of water dissociation on the surface. For instance, previous DFT calculations [51] and X-ray photoelectron spectroscopy experiments [52] indicate that a mixed dissociated and undissociated water adsorption takes place with a reported relative extent of dissociated water of 0.25. Based on a molecular dynamics study using an ab initio-based deep neural network potential, other authors observe only a 0.056 fraction of dissociated water after 25 ns of simulation time [53]. In addition, adiabatic potential energy surface calculations show that, according to thermodynamics, dissociative H+OH fragments are expected to very easily recombine into a water molecule [54], making the decomposition of water on this surface a reversible process. Car-Parrinello molecular dynamics (CPMD) calculations indicate that the dissociation takes place only on oxygen-defective (partially reduced) surfaces [55]. Based on all this information, the molecular adsorption mode of water was considered in this work to investigate the competition with nucleotides for the adsorption on the surface."

It is important to note that this is a first and approximated assessment since, on one side, the difference in the cavitation free energies of the nucleotides in bulk water and at the anatase (101)/water interface and, on the other, the variation in the water-water interactions when the nucleotide moves from the bulk water to the interface have not been included

Using the same level of theory as for all previous calculations (B3LYP-D*), the adsorption energy for water is computed to be -0.95 and -0.90 eV at the coverage of 0.25 and 1.0 ML, respectively. These values are in good agreement with previous theoretical works [51,56-60] (see **Table S1**).

In order to compare the adsorption energies of the water and the oligonucleotides we first consider the case of dihydrogen phosphate anion, as described in **Section 3.1.** We observe that the adsorption energy for the most stable monodented conformation is -1.02 eV, while for the most stable bidented conformation it is -2.01 eV (i.e. 1.005 eV per $Ti_{5c}$). These adsorption energies are slightly higher than those for water (-0.95 and -0.90 eV for at low and full coverage, respectively), suggesting that the dihydrogen phosphate anion is slightly favored in the competition.

Next, we consider a whole nucleotide, as described in **Section 3.3**. Considering the monodentate adsorption mode, we observe that two $Ti_{5c}$ surface atoms become involved. One of them interacts with the phosphate anion and the other with the nucleobase (depending on the nucleobase the interaction may be due to a carbonyl or to an aromatic N atom, see left panels in **Figure 5**). The adsorption energies for these configurations are -2.61, -3.37, -2.62, -2.04 eV for dAMP, dGMP, dCMP and dTMP, respectively. After dividing by the number of $Ti_{5c}$ (two) the values become: -1.305, -1.685, -1.31 and 1.02 eV, which must be compared with the water adsorption energy values (-0.95 and -0.90 eV for at low and full coverage, respectively). The conclusion is that, for dAMP, dGMP and dCMP, an eventual competition for the adsorption would be clearly in favor of the nucleotide.

Considering the dinucleotides in the most stable configuration (DOWN), four $Ti_{5C}$ surface atoms are involve in the adsorption (see **Figures 6** and **S2**). It means that values in **Table 2** normalized *per* number of $Ti_{5c}$ become: -1.32, -1.36, -1.33 and -1.15 eV for 2×dAMP, 2×dGMP, 2×dCMP and 2×dTMP, respectively. Also, here we observe that the eventual competition with water is in favor of the first three dimers.

As we mentioned above, this analysis is purely based on the adsorption energy value of the nucleotides vs those of water molecules at different coverages on the (101) surface. In a realistic situation, more complex processes can take place, such as the solvation of the oligonucleotide, the diffusion of water and nucleotide molecules to the surface, the interaction between oligonucleotide fragments, the interaction between water molecules at the interface, etc. These processes depend on external factors, such as pH, salt concentration and temperature, among others. Therefore, to fully understand and control the solvent competition, further studies using molecular dynamics techniques in a real aqueous environment should be performed. We are currently working on these topics and the results will be the subject of a future study.

## 4. Conclusions

In this work we have investigated the adsorption of nucleotides on the anatase (101) $TiO_2$ surface using hybrid density functional theory calculations including dispersion forces (B3LYP-D*). A thorough study was performed starting from the adsorption of an isolated phosphate group and of the four different nucleobases on the surfaces, to continue with an entire nucleotide, made of different bases, up to dinucleotide models.

We found that the phosphate can adsorb in four different ways: two monodentate and two bidentate, being the bidentate undissociated the most stable configuration. For the nucleobases, we found that the interaction with the surface can be established through of the carbonyl oxygen atom and/or through N atoms (either form the amine groups or the aromatic ones) bonded to a $Ti_{5c}$ surface atoms and through the H-bond with the $O_{2c}$ surface atoms. The adsorption strength for the bases is in the order $T < C < A < G$. The types of interactions between the nucleobases and the $TiO_2$ surface identified in this work are similar to those reported for other mineral or oxide surfaces, ranging from H-bonds to covalent binding between the aromatic N atoms or the O carbonyl atoms of the nucleobase and the uncoordinated metal atoms on the surfaces

When nucleotides were put on the surface, we determined three different ways in which they bind. In all cases, both the phosphate anion and the nucleobase are found to interact with the surface, in a collaborative adsorption. The resulting adsorption energies of nucleotides on the surface are close to the sum of the binding energy values obtained for the phosphate and nucleobase entities, when separately investigated. For dinucleotides, we studied the upward and the downward configurations (UP and DOWN), where we observed a collaborative adsorption. However, since in the DOWN configuration both the two phosphates and the two bases interact with the surface, the adsorption energy is much larger. Regarding to the optical properties, after adsorption of the nucleotide dimers on the slab model, the $\Delta_{HOMO-LUMO}$ gaps vary between 2.19 and 2.87 eV, which indicates an expected sensible red-shift of the band gap of the $TiO_2$ surface (4.26 eV) resulting from the DNA adsorption.

Finally, we considered the adsorption of competing water molecules at different coverage densities. This allows assessing on the stability of the nucleotides in aqueous media. We found out that, in general, the nucleotides and dinucleotides bind more strongly than water (mostly those containing adenine, guanine and cytosine), which, therefore, cannot cause their desorption. Further work will be performed in the future in order to accurately model the more complex real environment of longer oligonucleotides coated TiO$_2$ nanoparticles by considering the effect of temperature, pressure and pH.

## Acknowledgments

The authors are grateful to Lorenzo Ferraro for his technical help. The project has received funding from the European Research Council (ERC) under the European Union's HORIZON2020 research and innovation programme (ERC Grant Agreement No [647020]).

# Supplementary material

**Binding Groups of Oligonucleotides on TiO$_2$ Surfaces: phosphate anions or nucleobases?**

Federico A. Soria, Cristiana Di Valentin[*]

Dipartimento di Scienza dei Materiali, Università di Milano Bicocca,

via R. Cozzi 55 20125 Milano Italy

---

[*] Corresponding author: cristiana.divalentin@unimib.it

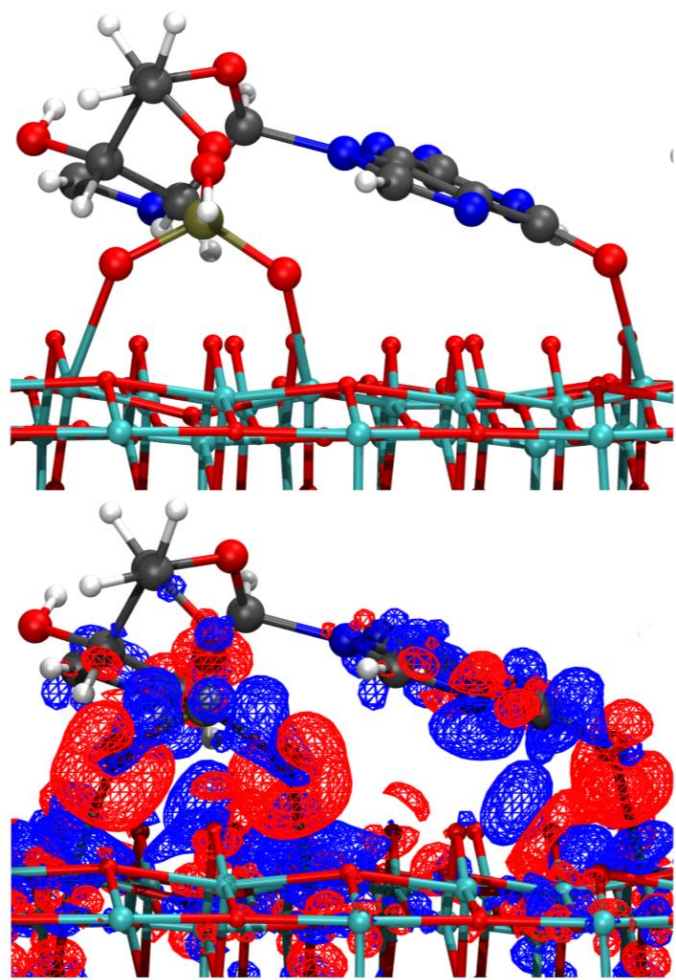

**Figure S1.** Charge density difference plot for dGMP (in a bidented undissociated mode) adsorbed on the TiO$_2$ (101) anatase surface. The red distribution corresponds to charge accumulation and blue correspond to depletion. The isosurface density value used is $2 \times 10^{-3}$ e/Å$^3$.

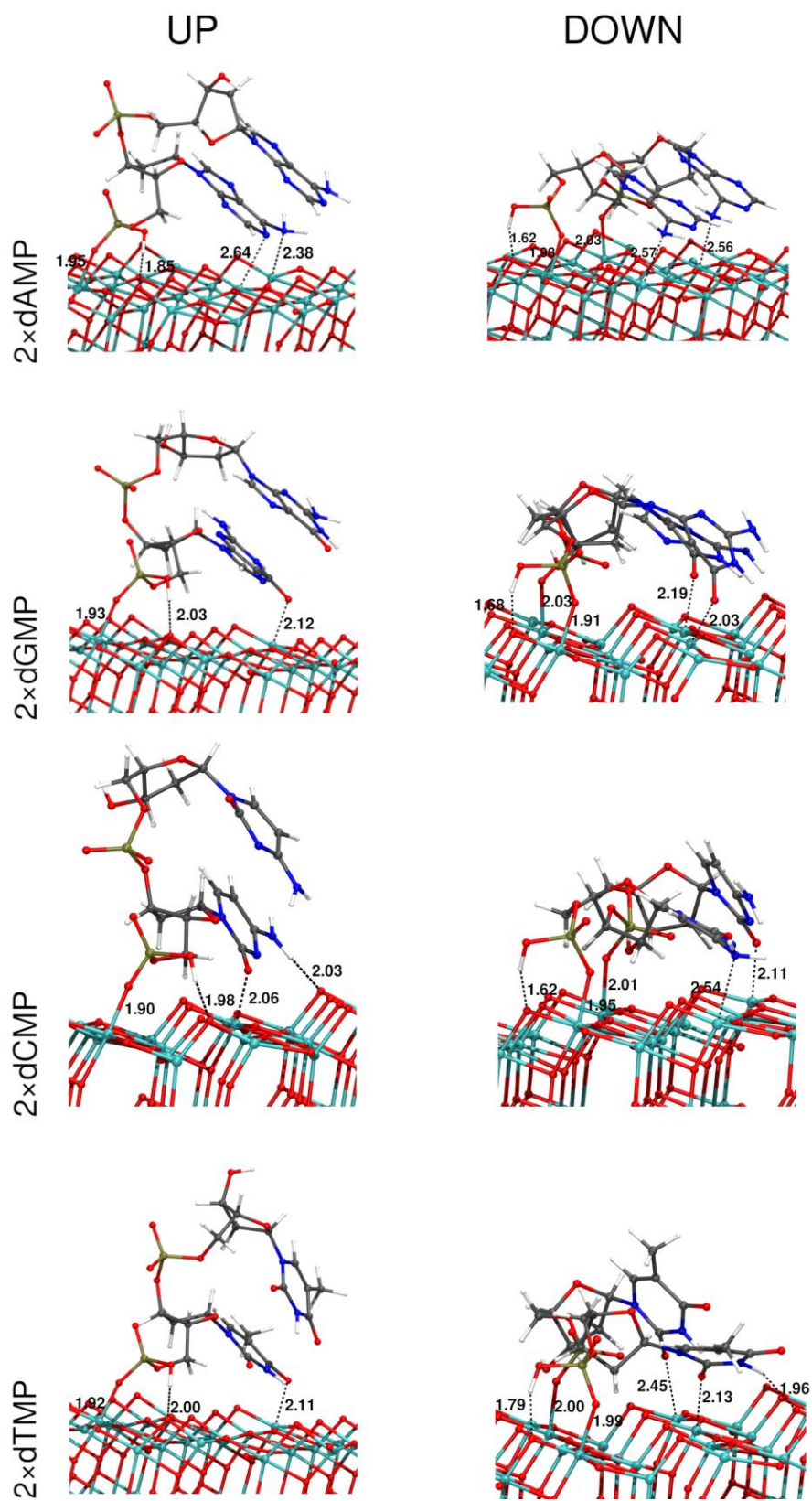

**Figure S2.** Optimized structures of the adsorption of a dimer of dAMP, dGMP, dCMP and dTMP on the TiO$_2$ (101) anatase surface in two different adsorption modes (see text) as obtained by B3LYP-D*. Distances in Å.

**Table S1.** Adsorption energies (expressed in eV) per water molecule adsorbed on the TiO$_2$ (101) anatase surface calculated with different DFT functionals.

| | coverage | $E_{ads}^{mol}$ |
|---|---|---|
| B3LYP-D*[present work] | 0.25 | -0.95 |
| | 1 | -0.90 |
| PBE0-D3 [1] | 0.25 | -1.03 |
| | 1 | -0.84 |
| PBE0 [1] | 0.25 | -0.78 |
| | 1 | -0.61 |
| HSE [2] | 0.25 | -0.84 |
| | 1 | -0.77 |
| PBE [2] | 0.25 | -0.71 |
| | 1 | -0.68 |
| PBE [3] | 0.25 | -0.67 |
| | 1 | -0.62 |
| PBE [4] | 0.25 | –0.74 |
| | 1 | –0.72 |
| PBE-D2 [3] | 0.25 | -0.91 |
| | 1 | -0.86 |
| PBE-D3 [2] | 0.25 | -0.93 |